\documentclass[aps,pra,preprint,superscriptaddress,longbibliography]{revtex4-2}

\usepackage{amsmath}
\usepackage{amsthm}
\usepackage{amsfonts}
\usepackage{amssymb}
\usepackage{xcolor,graphicx}
\usepackage{color}
\usepackage[pdftex]{hyperref} 
\usepackage{array}
\usepackage{dsfont}

\begin{document}
	
	\title{Noisy Cyclic Quantum Random Walk}
	
	\author{G. Juarez Rangel}
	\email[e-mail: ]{a01570628@tec.mx}
	\affiliation{Tecnologico de Monterrey, Escuela de Ingenier\'ia y Ciencias, Ave. Eugenio Garza Sada 2501, Monterrey, N.L., Mexico, 64849}	
	
	\author{B. M. Rodr\'iguez-Lara}
    \email[e-mail: ]{bmlara@upp.edu.mx, blas.rodriguez@gmail.com}
	\affiliation{Universidad Polit\'ecnica de Pachuca, Carr. Pachuca-Cd. Sahag\'un Km.20, Ex-Hda. Santa B\'arbara, Zempoala, 43830, Hidalgo, Mexico}	
	
	\date{\today}
	
\begin{abstract}
    We explore static noise in a discrete quantum random walk over a homogeneous cyclic graph, focusing on the spectral and dynamical properties of the system. 
    Using a three-parameter unitary coin, we control the spectral structure of the noiseless step operator on the unit circle in the complex plane.
    One parameter governs the probability amplitudes and induces two spectral bands with a gap proportional to its value.
    The half-sum of the two phase parameters rotates the spectrum and induces twofold degeneracy under specific conditions. 
    Degenerate spectra yield eigenstates with sinusoidal probability distributions, while non-degenerate spectra produce flat distributions.
    We introduce static phase noise in the sites and analyze its impact across two distinct propagation regimes.
    In the walk-on-the-line regime, which precedes a full traversal of the graph, we extract the spreading exponent~$\beta$ from the step-resolved mean squared displacement.
    We find that low participation ratios correlate with sub-diffusive spread, while high ratios correspond to ballistic or super-diffusive evolution.
    After the walker completes a cycle, finite-size effects dominate.
    In this walk-on-the-cycle regime, the spreading exponent no longer characterizes the behavior.
    We quantify localization using a convergence criterion based on the coefficient of variation of the mean squared displacement.
    Across both regimes, we identify a sharp crossover near static site noise strength $\phi_s = \pi/3$.
    This transition coincides with a drop in the participation ratio, a transition from diffusive to sub-diffusive spread in the walk-on-the-line regime, and a reduction in the saturation level of the mean squared displacement in the walk-on-the-cycle regime.
    Our results demonstrate that the eigenstate participation ratio provides a computationally efficient spectral diagnostic that anticipates localization across both regimes, offering an alternative to full dynamical simulations.
\end{abstract}

	
	\maketitle
	\newpage

\section{Introduction}

Anderson localization, first identified in disordered electronic systems, describes the inhibition of wave scattering caused by disorder in a medium \cite{Lahini2008p013906, Crespi2013p322328, Segev2013p197204, Vakulchyk2017p144204, Weidemann2021p576581, Longhi2023p2200658}. 
This phenomenon arises from the interference of multiple propagation paths and occurs in various physical systems, including optical, acoustic, and quantum waves \cite{Condat1987p6782, Weaver1990p0165, Billy2008p891, Piraud2014p063639, Yangtao2015p155402}. 
Random walks are an example of Anderson localization in the quantum domain, where the interplay between quantum coherence and randomness leads to diverse dynamical behaviors.
These dynamics have potential applications in areas such as quantum transport modeling \cite{Schreiber2011p180403, Nosrati2021p25702578, Buarque2023p064139, Almeida2024p064151} and quantum computing \cite{Derevyanko2018p1795, Wang2018p11361140}.

We are interested in Anderson localization in cyclic quantum random walks, where the walker is confined to a structure with periodic boundary conditions.
These conditions introduce unique constraints, affecting the onset and nature of localization in the presence of disorder. 
Here, we systematically explore how disorder influences the localization properties, analyzing the probability distribution over time, the inverse participation ratio, and the effects of static noise in the coin and site phases.
Our results reveal distinct dynamics of quantum localization in cyclic systems, offering insight into potential applications where generation and control of localized quantum states is crucial \cite{Rebentrost2009p033003, Sapienza2010p13521355, Hideaki2011p195139, Asch2012p194205, Bardarson2012p017202, Agarwal2015p160401, Chandrashekar2015p10005, Jonathan2015p104202, Wang2018p11361140, Lin2022p3229, Songyang2022p116801}. 
Additionally, we compare our findings with those in non-cyclic systems to highlight the distinct features of cyclic quantum walks.
Our analysis shows that cyclic topology can either mitigate or intensify localization effects depending on the disorder characteristics and system parameters.

The article is structured as follows. 
In Section~\ref{sec:S2}, we present the theoretical model of the discrete-time quantum random walk on a cyclic graph, introduce the three-parameter coin, and analyze its spectral effects. 
In Section~\ref{sec:S3}, we investigate how the coin parameters shape the eigenstates of the step operator, control the walker’s probability distribution, and influence the participation ratio as a spectral indicator of localization. 
In Section~\ref{sec:S4}, we introduce static phase noise in the sites and analyze its effect on the spreading behavior of the walker using the mean squared displacement and spreading exponent. 
To distinguish the influence of cyclic topology, we separate the results into two propagation regimes.
The walk-on-the-line regime, where the walker behaves as if on an unbounded line, Sec. \ref{sec:S5}, and the walk-on-the-cycle regime, where finite-size effects become dominate, Sec. \ref{sec:S6}. 
Section~\ref{sec:S7} proposes a gate-based quantum circuit implementation of our model.
We conclude in Section~\ref{sec:S8} with a summary of our findings and their relevance to spectral diagnostics and localization in noisy quantum walks.

\section{Quantum Random Walk in a Cyclic Graph} 
\label{sec:S2}

An ideal, lossless, discrete-time quantum random walk relies on two components, a coin and an adjoint pair of step operators \cite{Brun2003p032304}. 
The state of the coin defines the direction of the step. 
The step-coin operator produces a unitary process that preserves the total probability of the system in the ideal, lossless case.

We define a general unitary coin with three real parameters \cite{Vakulchyk2017p144204},
\begin{eqnarray}
    \hat{C}(\gamma, \theta, \phi) = \cos \gamma \vert 0 \rangle \langle 0 \vert + e^{i\theta} \sin \gamma \vert 0 \rangle \langle 1 \vert + e^{i\phi} \sin \gamma \vert 1 \rangle \langle 0 \vert - e^{i(\theta+ \phi)} \cos \gamma  \vert 1 \rangle \langle 1 \vert,
\end{eqnarray}
where the parameter $\gamma \in \left[ 0, \pi/2 \right]$ controls the probability amplitudes between the basis states and modifies the hopping strength, while the parameters $\theta, \phi \in \left[ 0, 2 \pi \right]$ control the coherence and interference in the quantum walk.
This coin is non-Hermitian and its eigenstates and eigenvalues are complex,
\begin{eqnarray}
        \vert c_{j} \rangle =&~ \alpha_{j} \,\vert 0 \rangle + \sqrt{ 1 - \vert \alpha_{j} \vert^{2}} \, \vert 1 \rangle, \nonumber \\
        c_{j} =&~ s + (-1)^{j} \sqrt{ e^{i \left( \theta + \phi \right)} + s^{2} },
\end{eqnarray}
with $s= \{ \left[ 1 - e^{i \left(\theta + \phi\right )} \right] \cos \gamma \}/2$ and $\alpha_{j} = \tilde{\alpha}_{j} / \sqrt{4 + \vert \tilde{\alpha}_{1 + (-1)^{j}}\vert^{2}}$, where $\tilde{\alpha}_{j} = 2  ( e^{-i \phi} c_{j} + e^{i \theta} \cos \gamma ) \csc^{1-j} \gamma$ and $j=0,1$.
For example, the parameter set $\{ \gamma, \theta, \phi \} = \{\pi/4, 0, 0\}$ yields the Hadamard coin, which produces a balanced superposition of the walker position states. 
Applied to a localized initial state $\vert j \rangle \vert +\rangle$, the Hadamard coin tends to asymmetrically distribute the probability amplitude  along the line \cite{Nayak2012p, Chandrashekar2007p022316}. 
In contrast, the parameter set $\{\gamma, \theta, \phi\} = \{\pi/4, \pi/2, \pi/2\}$ yields the symmetric coin, which produces a balanced superposition of the walker's position. 
Applied to a localized initial state $\vert j \rangle \vert + \rangle$, this coin symmetrically distributes the probability amplitude equally along the line. 

\begin{figure}
    \includegraphics[scale=1]{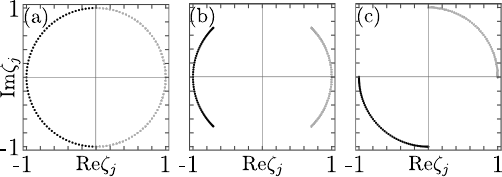}
    \caption{Eigenvalues $\zeta_{k}$ in the complex plane for a cyclic graph with 128 identical sites and coin parameters $\left\{ \gamma , \left( \theta + \phi\right)/2 \right)$ equal to (a) $\left\{ 0, 0 \right\}$, (b) $\left\{ \pi/4, 0 \right\}$, and (c) $\left\{ \pi/4, \pi/4 \right\}$ with $\theta = \phi$.}
    \label{fig:Figure1}
\end{figure}

The step operator in the discrete cyclic quantum random walk,
\begin{equation}
    \hat{S}_{N} = \left( \hat{V}_{N} \otimes \hat{P}_{0}  + \hat{V}_{N}^{\dagger}\otimes \hat{P}_{1} \right) \left[ \hat{1} \otimes \hat{C}\left( \gamma, \theta, \phi \right) \right],
    \label{eq: StepOperator}
\end{equation}
is proportional to the clockwise and counter-clockwise step operators on the cyclic graph,
\begin{eqnarray}
        \hat{V}_{N} &= \sum_{j = 1}^{N} \vert \mathrm{mod}_{N} (j) + 1 \rangle \langle j \vert, \nonumber \\
        \hat{V}_{N}^{\dagger} &= \sum_{j = 1}^{N} \vert \mathrm{mod}_{N} (j - 2) + 1 \rangle \langle j \vert.
\end{eqnarray}
where we used $\mathrm{mod}_{N} ( - 1) = N - 1$. 
The step operators move the state of the system to the nearest neighbor in a clockwise or counter-clockwise direction, conditional on the state of the coin after the coin projector operator $\hat{P}_{j} = \vert j \rangle \langle j \vert$ acts on it.
By moving into a reference frame rotating at the site frequency and using the quantum Fourier transform \cite{Weinstein2001p18891891, Stevens2018pA31A37, Jaramillo2019p515}, 
\begin{eqnarray}
    \hat{F}_{N} = \frac{1}{\sqrt{N}} \sum_{j,k=1}^{N} e^{i\frac{2\pi}{N}((j-1) ( k-1) } \vert j \rangle \langle k \vert,
\end{eqnarray}
we diagonalize the step operators,
\begin{eqnarray}
        \hat{F}_{N}^{\dagger} \hat{V}_{N} \hat{F}_{N} =&~ \hat{\Lambda}_{N},  \nonumber \\ \hat{F}_{N}^{\dagger} \hat{V}_{N}^{\dagger} \hat{F}_{N} = &~\hat{\Lambda}_{N}^{\dagger},
\end{eqnarray}
in terms of the diagonal clock operator \cite{Jozef2006p0223}, 
\begin{eqnarray}
\hat{\Lambda}_{N} = \sum_{j=1}^{N} e^{-i\frac{2\pi}{N}(j-1)} \vert j \rangle \langle j \vert,
\end{eqnarray}
providing the eigenstates and eigenvalues \cite{Stevens2018pA31A37}, 
\begin{eqnarray}
        \vert \lambda_{k} \rangle =&~ \hat{F}_{N}|k\rangle  = \sum_{j=1}^{N} e^{i \frac{2 \pi}{N} (j-1) (k-1)} \vert j \rangle, \nonumber \\
        \lambda_{k} =&~  e^{-i\frac{2\pi}{N}(k-1)},
\end{eqnarray}
distributed nonlinearly on a line with $k=1,2, \ldots, N$. 
We now use the eigenstates of the coin and the step operators on the cyclic graph to build those of the discrete step operator,
\begin{eqnarray}
        \vert \zeta_{k} \rangle =&~ \cos \vartheta \vert \lambda_{k}, 0 \rangle + e^{i \varphi_{-} } \sin \vartheta |\lambda_{k}, 1\rangle, \nonumber \\
        \zeta_{k} =&~ e^{ -i \left\{ \arcsin \left[ \cos \gamma \sin \varphi_{+} \right] - \frac{\theta + \phi}{2} \right\} },
\end{eqnarray}
with real angles $\vartheta = - \arctan \left[ \left( \cos \varphi_{+} \pm \sqrt{ \tan^{2} \gamma + \cos^{2} \varphi_{+} } \right) /\tan^{2} \gamma \right]$ and $\varphi_{\pm} = 2 \pi k / N \pm \left( \theta \pm \phi \right)/2$. 
The eigenvalues depend only on the half-sum of the two coin phases $\left( \theta + \phi \right)/2$ rather than their individual values. 
The argument of the eigenvalues, $\omega_{k} = \arg{\zeta_{k}}$, forms two bands $\left( \theta + \phi \right)/2 + \gamma \mp \pi/2  < \omega_{k} < \left( \theta + \phi \right)/2  - \gamma \pm \pi/2$, which close for $\gamma = 0$. 
In the complex plane, the $\zeta_{k}$ eigenvalues trace two $(\pi - 2 \gamma)$ arcs of a unit circle, rotated by the half-sum of the coin phases relative to the real axis, Fig. \ref{fig:Figure1}.

\begin{figure}
    \includegraphics[scale=1]{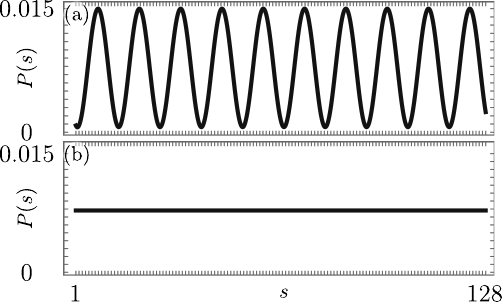}
    \caption{Probability distribution on the cyclic graph sites $P(s)$, after tracing out the coin, for (a) the first eigenstate in the spectral degenerate case with $\left( \theta + \phi \right)/2 = m \pi / N$ and $m=64$ and $N=128$, and (b) the non-degenerate case with $\left( \theta + \phi \right)/2 = 5 \pi / 14$.}
    \label{fig:Figure2}
\end{figure}

Pairs of degenerate eigenvalues appear when the half-sum of the coin phases satisfies $\left( \theta + \phi \right)/2 = m \pi / N$ with $m=0, 1, \ldots, N-1$. 
Otherwise, no spectral degeneracy exists. 
In the degenerate case, the probability distribution for the step eigenfunctions, after tracing out the coin, forms a sinusoidal pattern shape with maximum values of $2/N$, Fig. \ref{fig:Figure2}(a). 
In contrast, for the non-degenerate case, the distribution is flat with value $1/N$, Fig. \ref{fig:Figure2}(b).

\section{Effect of the Coin}
\label{sec:S3}

To predict the behavior of the walker under a coin, we use the participation ratio,
\begin{eqnarray}
    \text{PR}( \zeta_{j} ) = \frac{\left( \sum_{s=1}^{N} P(s) \right)^2}{\sum_{s=1}^{N} P^{2}(s)},
\end{eqnarray}
defined in terms of the probability $P(s)$ of finding the step eigenstate $\vert \zeta_{j} \rangle$ at site $s$. 
The participation ratio ranges from one to the number of sites $N$. 
It is a spectral quantity that measures the spatial localization of the eigenstates of the step operator.  

Low participation ratios indicate that most eigenstates are localized, causing the walker to concentrate around the initial position. 
High participation ratios suggest delocalization, where the walker is likely to spread out from the initial position.
In this sense, we use it as a static diagnostic tool to anticipate localization or delocalization.

\begin{figure}
    \includegraphics[scale=1]{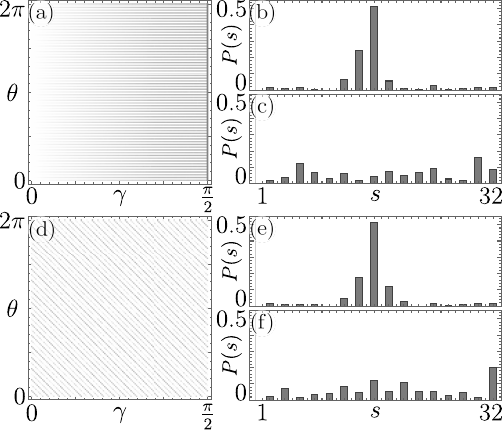}
    \caption{Average participation ratio for the step eigenstates using a coin with (a) variable $\gamma$ and $\theta=\phi$, and (d) fixed $\gamma = \pi/4$ with variable $\theta$ and $\phi$. 
    White corresponds to a $\text{PR}$ equal to the size of the graph with $N=32$. 
    Final state probability distribution on the cyclic graph sites $P(s,m)$, after tracing out the coin, for $N=32$, $m=1000$, and coin parameters $\{ \gamma, \theta, \phi\}$ producing localization, (b) $\left\{1/4, 5/32, 5/32 \right\} \pi$ and (e) $\left\{1/4, 13/32, 17/32 \right\}$, as well as delocalization, (c) $\left\{ 1/4, 5.5/32, 5.5/32 \right\} \pi$ and (f) $\left\{ 1/4, 12/32, 17/32 \right\} \pi$.}
    \label{fig:StaticCoin}
\end{figure}

Fig. \ref{fig:StaticCoin}(a) shows the mean participation ratios for a coin with variable parameters $\gamma$ and $\theta=\phi$ on a cyclic graph with $N=32$ sites. 
We use an initial state localized at site $s_{0}=16$ to analyze the final state after $m=1000$ steps. 
Quantum walks with a mean participation ratio of $2.99$ localize the walker, Fig. \ref{fig:StaticCoin}(b), while those with a mean participation ratio of $11.60$ show delocalization, Fig. \ref{fig:StaticCoin}(c). 
Figs.  \ref{fig:StaticCoin}(d)-(f) present equivalent results for fixed $\gamma = \pi/4$ with variable $\theta$ and $\phi$.
In Fig. \ref{fig:StaticCoin}(b) and Fig. \ref{fig:StaticCoin}(e), we focus on the second minimum mean participation ratio, observing minimal spreading. 
Choosing the absolute minimum, $\text{PR} = 2$, would merely shift the initial state left and right of its original position. 
We optimized these results for cyclic graphs with $2^{n}$ sites, where $n = 5, 6, 7, 8$, consistently obtaining similar outcomes across each realization.
However, visualizing the features becomes harder for larger graphs.

\section{Static Noise in the Sites}
\label{sec:S4}

Phase noise introduced in the cyclic graph step operators is equivalent to altering the self-energy terms in the continuous-time Hamiltonian. 
As a result, local perturbations can either enhance or inhibit interference, affecting localization and diffusion properties.
We consider a diagonal noise operator,
\begin{eqnarray}
    \hat{D}_{\mathrm{noise}}(\varphi) = \sum_{s=1}^{N} e^{i \phi_{s} } \vert s \rangle \langle s \vert,
\end{eqnarray}
such that the noisy step operator is given by
\begin{eqnarray}
    \hat{S}_{\mathrm{noise}}(\varphi) = \left( \hat{D}_{\mathrm{noise}}(\varphi) \otimes \hat{1} \right) \hat{S}_{N},
\end{eqnarray}
with real random phase noise in the range $\phi_{s} \in \left[ -\varphi, \varphi \right]$. 
In the following, we optimize the random phase range to produce localization, using the participation ratio for the eigenvalues of the noisy step operator.

To quantify the spread of the walker at each step, we use the mean squared displacement on the cycle,
\begin{eqnarray}
\Delta x^{2}(m) = \sum_{s=1}^{N} \left[ \Delta x(m)\right]^{2} P(s,m),
\end{eqnarray}
in terms of the probability $P(s,m)$ of finding the walker at site $s$ after $m$ steps, and the mean displacement,
\begin{eqnarray}
    \Delta x(m) = \mathrm{min}\left[ \vert x(m) - x(0) \vert, N - \vert x(m) - x(0) \vert \right],
\end{eqnarray}
which is the minimum distance between the walker mean position at step $m$ and its initial site. 
Here, the mean position of the walker on the cyclic graph is $x(m) = \langle \psi(m) \vert \hat{n} \vert \psi(m) \rangle$, where the site number operator is $\hat{n} = \sum_{s=1}^{N} s \vert s \rangle \langle s \vert$.
Fitting the mean squared displacement to a power law,
\begin{eqnarray}
    \Delta x^{2}(m) \approx \alpha m^{\beta},
\end{eqnarray}
allows us to quantify how fast the walker spreads over each step \cite{Nosrati2021p25702578, Sassetti1996p2914, Almeida2024p064151}. 
The rate of spread~$\alpha$ quantifies how fast the displacement grows, but it does not determine the spreading regime. 
The spreading exponent~$\beta$ is the relevant dynamical quantity, characterizing how the walker spreads as a result of the evolution. 
This contrasts with the participation ratio, a static spectral diagnostic computed from the eigenstates of the step operator, independent of the walker's actual trajectory.

The power-law fit we use to extract $\beta$ is only meaningful in the regime where the quantum walk resembles a walk on a line, before the walker completes a full traversal of the cycle. 
Beyond this regime, the spreading exponent $\beta$ no longer captures the asymptotic behavior. 
In our simulations, we restrict the fitting range to the initial steps where the system resembles a quantum walk on a line with open boundaries.
If $\beta = 2$, the walker spreads ballistically; $\beta = 1$ indicates diffusive spread, while $\beta \in (1,2)$ corresponds to super-diffusive spread.
For $\beta \in (0,1)$, the walker exhibits sub-diffusive spread.

The quantum walk on a cyclic graph exhibits qualitatively different behavior depending on whether the walker has explored the full extent of the cycle. 
To account for this, we distinguish between two regimes in our analysis. 
The walk-on-the-line regime is restricted to a finite number of steps during which the walker has not yet completed a full traversal of the graph. 
In this regime, the dynamics are equivalent to those of a quantum walk on an unbounded line, and the effects of the cyclic topology are null. 
In contrast, the walk-on-the-cycle regime covers a larger number of steps, where the walker explores the entire graph and begins to experience finite-size effects due to the periodic boundary conditions.

\section{Walk-on-the-line regime}
\label{sec:S5}

We analyze the walker dynamics in the walk-on-the-line regime. 
This regime is restricted to a limited number of steps before the walker completes a full traversal of the cycle, and the effects of the periodic boundary conditions become relevant. 
Within this regime, we extract the spreading exponent~$\beta$ from a power-law fit to the mean squared displacement. 
This also provides a consistent setting to connect the static participation ratio, computed from the spectrum of the step operator, with the dynamical behavior of the walker.

\begin{figure}
    \includegraphics[scale=1]{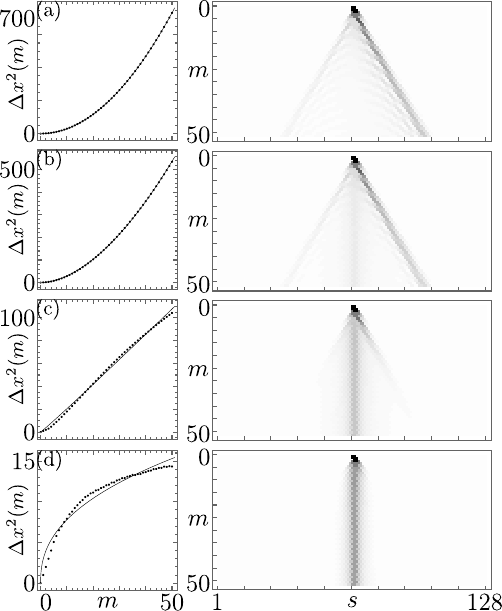}
    \caption{Mean squared displacement $\Delta x^{2}(m)$ (left column) and probability of finding the walker at site $s$ at step $m$, $P(s,m)$, for a graph with size $N = 128$ with coin parameters $\left\{\pi/4, 0, 0\right\}$ for noise levels $\phi_{s}$, rate of spread $\alpha$, and spreading exponent $\beta$. 
    The values $\left\{ \phi_{s}, \alpha, \beta \right\}$ are equal to (a) $\left\{ 0, 0.297, 1.996 \right\}$, (b) $\left\{ \pi/10, 0.480, 1.794 \right\}$, (c) $\left\{ \pi/3, 1.901, 1.032 \right\}$, (d) $\left\{ \pi , 2.973, 0.419 \right\}$.}
    \label{fig:HadCoinProp}
\end{figure}

Figure \ref{fig:HadCoinProp} presents the mean squared displacement and probability distribution of a walker using a Hadamard coin on a cyclic graph with $N=128$ sites. 
The plot focuses on the initial steps, where the system resembles a quantum walk on a line.
The cases with higher mean participation ratios, shown in Fig.~\ref{fig:StaticCoin}(c) and Fig.~\ref{fig:StaticCoin}(f), lead to ballistic and super-diffusive spread, Fig.~\ref{fig:HadCoinProp}(a) with $\beta =1.996$ and Fig.~\ref{fig:HadCoinProp}(b) with $\beta = 1.794$. 
Fig.~\ref{fig:HadCoinProp}(c) shows diffusive spread with $\beta = 1.032$.
The cases with lower mean participation ratios, shown in Fig.~\ref{fig:StaticCoin}(b) and Fig.~\ref{fig:StaticCoin}(e), result in sub-diffusive spread, Fig.~\ref{fig:HadCoinProp}(d) with $\beta = 0.419$. 
A noise level of $\phi_{s} = \pi/3$ marks the threshold between sub-diffusive and super-diffusive spread.

\begin{figure}
    \includegraphics[scale=1]{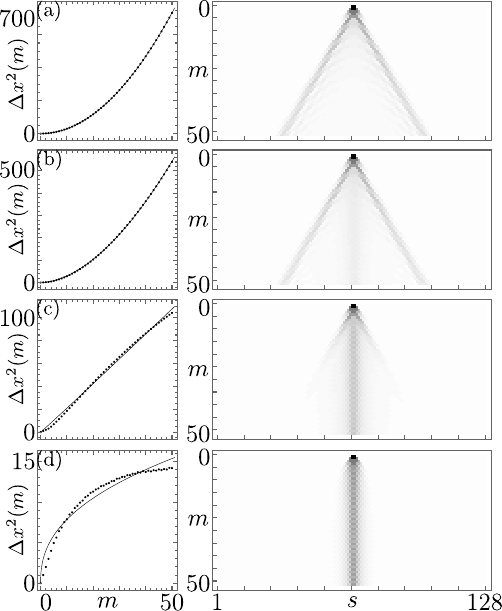}
    \caption{Mean squared displacement $\Delta x^{2}(m)$ (left column) and probability of finding the walker at site $s$ at step $m$, $P(s,m)$, for a graph with size $N = 128$ with coin parameters $\left\{\pi/4, \pi/2, \pi/2\right\}$ for noise levels $\phi_{s}$, rate of spread $\alpha$, and spreading exponent $\beta$. 
    The values $\left\{ \phi_{s}, \alpha, \beta \right\}$ are equal to (a) $\left\{0, 0.298, 1.996 \right\}$, (b) $\left\{ \pi/10, 0.475, 1.800 \right\}$, (c) $\left\{ \pi/3, 2.005, 1.019 \right\}$, (d) $\left\{ \pi , 3.006, 0.417 \right\}$.}
    \label{fig:SymCoinProp}
\end{figure}

Figure~\ref{fig:SymCoinProp} shows the mean squared displacement and the probability distribution of the walker for a symmetric coin, focusing on the initial steps, where the walk exhibits linear quantum behavior. 
The results are comparable to those of the Hadamard coin, except that the symmetric coin yields symmetric probability distributions.

\begin{figure}
    \centering
    \includegraphics[scale = 1]{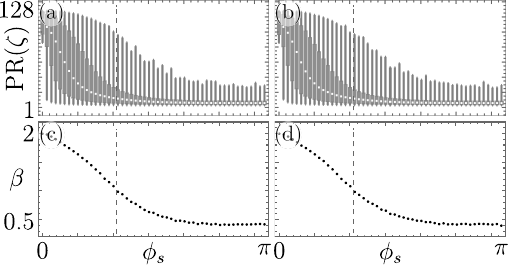}
    \caption{Participation ratio boxplots (first row) depicting the distribution and spreading exponent~$\beta$ for different noise levels $\phi_{s}$ with coin parameters (a),(c) $\left\{\pi/4, 0, 0\right\}$  and (b), (d) $\left\{\pi/4, \pi/2, \pi/2\right\}$. The vertical gray dashed line marks $\phi_s = \pi/3$, separating super-diffusive and sub-diffusive regimes.}
    \label{fig:PR_Alpha_VS_Noise}
\end{figure}

Figure~\ref{fig:PR_Alpha_VS_Noise} shows the participation ratio boxplot distribution and spreading exponent~$\beta$ at different noise levels for the Hadamard coin (left column) and the symmetric coin (right column). 
For both coins, we observe a direct relationship between the participation ratio and the spreading exponent~$\beta$, with a minimum value at the maximum noise level $\phi_{s} = \pi$.

We identify the noise level $\phi_s = \pi/3$ as a crossover threshold separating super-diffusive and sub-diffusive regimes. 
This value is not derived from an analytical expression, but obtained numerically from the behavior of the spreading exponent $\beta$. 
Across all cases analyzed, we observe that $\beta > 1$ for $\phi_s < \pi/3$, while $\beta < 1$ for $\phi_s > \pi/3$, marking the transition from delocalized to localized behavior.
This crossover is accompanied by a sharp drop in the participation ratio of the step operator’s eigenstates, reinforcing the connection between eigenstate localization and suppressed spreading. 
The trend persists for both Hadamard and symmetric coins and appears robust for graphs with diverse size $N$. 

\section{Walk-on-the-cycle regime}
\label{sec:S6}

After the walker completes a full traversal of the graph, the effects of periodic boundary conditions dominate the evolution. 
We refer to this large number of steps regime as the walk-on-the-cycle regime. 
Here, the mean squared displacement no longer follows a power law and cannot be reliably described by the spreading exponent $\beta$. 
Instead, we study the dynamical signatures of large number of steps and the onset of localization through direct observation of the mean squared displacement and its convergence.

\begin{figure}
    \includegraphics[scale=1]{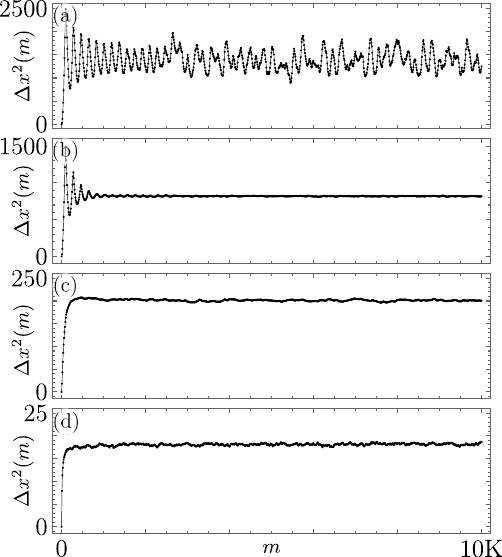}
    \caption{Mean squared displacement $\Delta x^{2}(m)$ over 10,000 steps for a cyclic graph with $N = 128$ using the Hadamard coin. Parameters are identical to those in Fig.~\ref{fig:HadCoinProp}, (a) $\left\{ 0, 0.297, 1.996 \right\}$, (b) $\left\{ \pi/10, 0.480, 1.794 \right\}$, (c) $\left\{ \pi/3, 1.901, 1.032 \right\}$, (d) $\left\{ \pi , 2.973, 0.419 \right\}$.}
    \label{fig:HadStaticCoinPropLarge}
\end{figure}

\begin{figure}
    \includegraphics[scale=1]{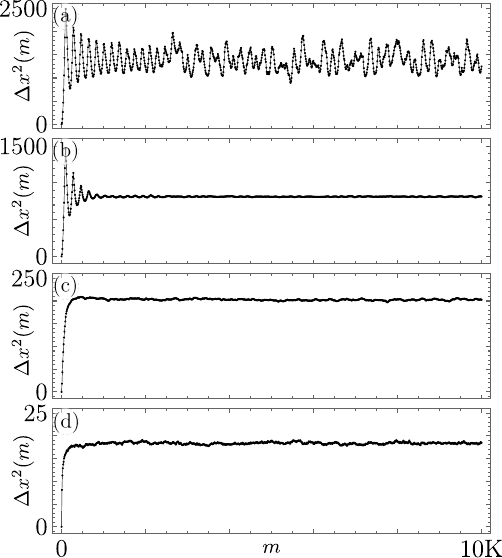}
    \caption{Mean squared displacement $\Delta x^{2}(m)$ over 10,000 steps for a cyclic graph with $N = 128$ using the symmetric coin. Parameters are identical to those in Fig.~\ref{fig:SymCoinProp}, (a) $\left\{0, 0.298, 1.996 \right\}$, (b) $\left\{ \pi/10, 0.475, 1.800 \right\}$, (c) $\left\{ \pi/3, 2.005, 1.019 \right\}$, (d) $\left\{ \pi , 3.006, 0.417 \right\}$.}
    \label{fig:SymStaticCoinPropLarge}
\end{figure}

Figure~\ref{fig:HadStaticCoinPropLarge} and Fig.~\ref{fig:SymStaticCoinPropLarge} show the mean squared displacement over 10,000 steps on a cyclic graph with $N = 128$ sites for the Hadamard and symmetric coins, respectively. 
The simulations run for 50,000 steps, but we restrict the plots to the first 10,000 for readability, as longer time windows offer no additional insight. 
To improve clarity without losing resolution, we record the mean squared displacement every 10 steps, yielding a representative view of the walker’s long-time evolution.

As the walk enters the walk-on-the-cycle regime, the behavior of the mean squared displacement changes qualitatively. 
The spreading exponent $\beta$, extracted from the initial step range, no longer describes the behavior once the walker has traversed the graph. 
In this regime, we characterize the dynamics by qualitative features of the step-dependent mean squared displacement. 
For the parameter sets associated with fast initial spread, the mean squared displacement exhibits persistent oscillations without stabilization, Fig.~\ref{fig:HadStaticCoinPropLarge}(a) and Fig.~\ref{fig:SymStaticCoinPropLarge}(a). 
At intermediate noise levels, the oscillations gradually diminish and the mean squared displacement approaches a stable value, Fig.~\ref{fig:HadStaticCoinPropLarge}(b) and Fig.~\ref{fig:SymStaticCoinPropLarge}(b). 
For higher noise levels, the walker reaches a steady displacement over fewer steps, indicating effective localization before completing additional traversals, Fig.~\ref{fig:HadStaticCoinPropLarge}(c,d) and Fig.~\ref{fig:SymStaticCoinPropLarge}(c,d).

To quantify long-time behavior, we use the coefficient of variation of the mean squared displacement as a convergence diagnostic. 
This dimensionless measure quantifies relative fluctuations over a moving window of size $n = \max\{10, N/4\}$ to avoid false convergence due to undersampling. 
We adopt the bias-corrected form from Ref.~\cite{Sokal1995p9},
\begin{equation}
\mathrm{CV} = \left(1 + \frac{1}{4n}\right) \frac{s}{\bar{x}},
\end{equation}
where $s$ and $\bar{x}$ are the standard deviation and mean of the mean squared displacement over the window. 
We define convergence as $\mathrm{CV} \le 0.01$ and use the corresponding $\bar{x}$ as the saturation level.

\begin{figure}
    \centering
    \includegraphics[scale = 1]{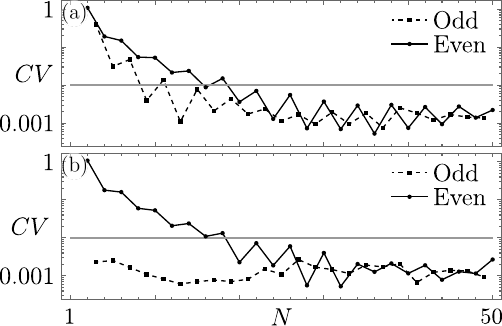}
    \caption{Minimum unbiased coefficient of variation over 5,000 steps at noise level $\phi_s = \pi$ with coin parameters (a) $\{\pi/4, 0, 0\}$ and (b) $\{\pi/4, \pi/2, \pi/2\}$, showing results for even and odd graph sizes.}
    \label{fig:MinCVvsN}
\end{figure}

Figure~\ref{fig:MinCVvsN} shows the minimum coefficient of variation within the first 5,000 steps at maximum noise level $\phi_s = \pi$ for different graph sizes. 
Panel (a) uses the Hadamard coin and panel (b) the symmetric coin. 
In both cases, convergence is more readily achieved for even $N$. 
For the Hadamard coin, convergence is inconsistent for small odd $N$, while for the symmetric coin, consistent convergence appears already at $N = 3$.

\begin{figure}
    \centering
    \includegraphics[scale = 1]{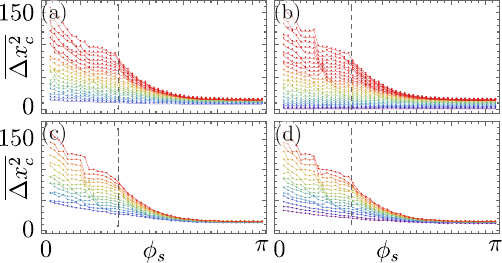}
    \caption{Saturation level of the mean squared displacement $\overline{\Delta x^2_c}$ as a function of noise strength $\phi_s$, for coin parameters (a)–(b) $\{\pi/4, 0, 0\}$ and (c)–(d) $\{\pi/4, \pi/2, \pi/2\}$. 
    The top row shows odd site numbers; the bottom row shows even ones. Colors represent the number of sites, from purple (fewest) to red (most), with site ranges: (a) 13–49, (b) and (d) 20–50, and (c) 3–49. 
    The vertical gray dashed line marks $\phi_s = \pi/3$, separating super-diffusive and sub-diffusive regimes.}
    \label{fig:MSDConvergence}
\end{figure}

Figure~\ref{fig:MSDConvergence} shows the saturation level of the mean squared displacement $\overline{\Delta x^2_c}$ as a function of the noise level $\phi_s$ for various graph sizes. 
Panels (a)–(b) correspond to the Hadamard coin and panels (c)–(d) to the symmetric coin; top rows show odd $N$, bottom rows even $N$. 
Colors encode graph size, with lighter colors for larger $N$. 
The vertical dashed line at $\phi_s = \pi/3$ marks the crossover from super-diffusive to sub-diffusive dynamics.

As the noise strength increases, the saturation level decreases, reflecting enhanced localization. 
For noise levels above the crossover $\phi_s = \pi/3$, the walker saturates at lower displacement values, provided the graph size is large enough. 
In this case, the walker effectively localizes before completing additional traversals of the graph. 
In contrast, below the crossover, saturation is either delayed or absent, indicating that the walker continues to explore the graph with no stable displacement level. 
Although this crossover was initially identified through the spreading exponent $\beta$, its significance persists beyond the walk-on-the-line regime and remains visible in the convergence trends at larger step counts.

\section{Quantum Circuit Implementation}
\label{sec:S7}

In gate-based quantum computing, a general unitary operator may be implemented using a sequence of three rotations, $\hat{R}_{z}(\alpha)\hat{R}_{y}(\theta)\hat{R}_{z}(\beta)$.
We define rotations around the Pauli-$j$ axis as $\hat{R}_{j}(\vartheta) = \cos(\vartheta/2)\hat{1} + i\sin(\vartheta/2)\hat{\sigma}_{j}$ with $j = x, y, z$ \cite{NielsenChuang2010}. 
Thus, the coin operator, 
\begin{eqnarray}
\hat{C}(\gamma, \theta, \phi) = \hat{R}_{z}(\phi)\hat{R}_{y}(2\gamma)\hat{R}_{z}(\theta + \pi),
\end{eqnarray}
may be implemented in such a quantum computer. 

\begin{figure}
\includegraphics[scale=1]{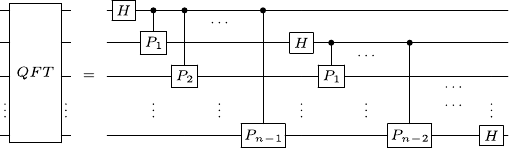}
\caption{Quantum Fourier transform (QFT) circuit for $n$ logical qubits, where $P_{j}$ is the phase gate with rotation angle $\pi/2^{j}$.}
\label{fig: QFT_Circuit}
\end{figure}

The step operator relies on the quantum Fourier transform (QFT), which scales quadratically with the number of qubits \cite{NielsenChuang2010}. 
Fig. \ref{fig: QFT_Circuit} illustrates the QFT algorithm for $n$ logical qubits. 
The QFT uses the phase gate $P_{j} = \vert 0 \rangle \langle 0 \vert + \vert 0 \rangle \langle 1 \vert + \vert 1 \rangle \langle 0 \vert + e^{i \pi / 2^{j}} \vert 1 \rangle\langle 1 \vert| $, with rotation angle $\pi / 2^{j}$, to achieve the desired diagonalization.

\begin{figure}
\includegraphics[scale=1]{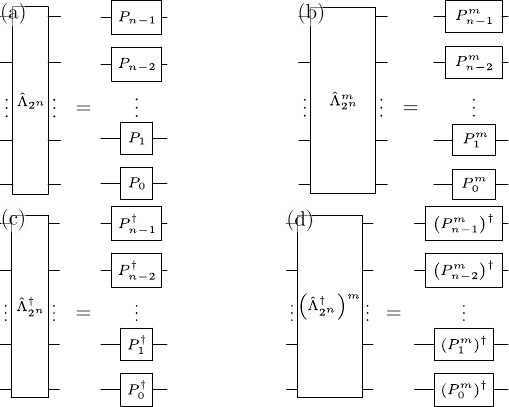}
\caption{Circuit implementations for the diagonal (a) clockwise and (c) counter-clockwise operators, and (b) and (d) their higher-order iterations , respectively. 
Here, $P_{j}$ and $P_{j}^{m}$ represent phase gates with rotation angles $\pi/2^{j}$ and $m\pi/2^{j}$, in that order.}
\label{fig: ClockOperator_Circuits}
\end{figure}

The diagonal clockwise and counter-clockwise operators can be implemented using controlled phase gates. 
Fig. \ref{fig: ClockOperator_Circuits} presents the corresponding circuits for $N = 2^{n}$ sites. 
These circuits exploit the additive property of phase gates, $P(\theta)P(\phi) = P(\theta+\phi)$, enabling efficient implementation of higher-order iterations.

\begin{figure}
\includegraphics[scale=1]{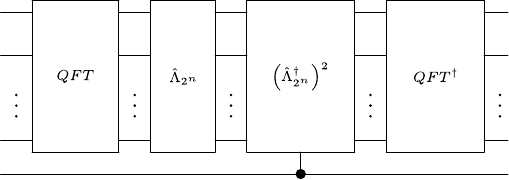}
\caption{Efficient step operator circuit for $n$ logical qubits, incorporating diagonal clockwise and counter-clockwise operators with phase gates $P_{j}$.}
\label{fig: StepOperator_Circuit}
\end{figure}

We show an optimized step operator circuit in Fig. \ref{fig: StepOperator_Circuit}. 
This approach reduces the reliance on two-qubit gates, which are typically more error-prone, by leveraging the diagonal representation of the clock operators. 

\begin{figure}
\includegraphics[scale=1]{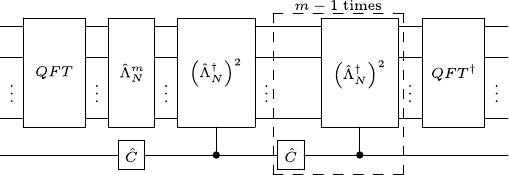}
\caption{Algorithm for performing multiple steps in the cyclic graph with coin operator $\hat{C}$.}
\label{fig: Multiple_Steps_Circuit}
\end{figure}

\begin{figure}
\includegraphics[scale=1]{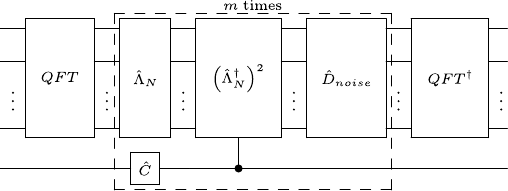}
\caption{Algorithm for performing multiple steps with static phase noise on the cyclic graph.}
\label{fig: Noisy_Steps_Circuit}
\end{figure}

For multiple-step implementations, the circuit in Fig. \ref{fig: Multiple_Steps_Circuit} implements repeated applications of the step operator. 
When introducing static noise into the sites, as described in Section IV, the noise operator must be incorporated into the Fourier basis. 
Fig. \ref{fig: Noisy_Steps_Circuit} demonstrates this integration. 
While the Fourier-transformed noise operator can be computationally intensive, efficient decomposition can be left to a transpiler, such as in the Qiskit framework.

\section{Conclusion}
\label{sec:S8}

We investigated Anderson localization in discrete quantum walks on cyclic graphs by introducing static phase disorder at the sites. 
In the absence of noise, the walker dynamics are controlled by the coin parameters, with the half-sum of the coin phases determining the participation ratio of the step operator eigenstates. 
This spectral diagnostic provides a static indicator of potential localization or delocalization.

To resolve the qualitative shift in behavior due to noise, we introduced two distinct propagation regimes. 
In the walk-on-the-line regime, the walker has not yet traversed the entire cycle, and the step-dependent mean squared displacement follows a power-law. 
We extracted the spreading exponent~$\beta$ as a dynamical observable to characterize the evolution, observing a transition from super-diffusive to sub-diffusive spread around a critical noise level $\phi_s = \pi/3$. 
This threshold was identified numerically and appears robust across coin types and system sizes.

In the walk-on-the-cycle regime, reached after the walker completes a full traversal, periodic boundary effects dominate and the spreading exponent no longer captures the evolution. 
Instead, we analyzed the saturation behavior of the mean squared displacement using a convergence criterion based on the coefficient of variation. 
For $\phi_s > \pi/3$, the walker localizes before completing another full traversal, and the mean squared displacement stabilizes at a noise-dependent value. 
In contrast, for $\phi_s < \pi/3$, no saturation occurs within the observed step range, and the walker continues to explore the graph through coherent interference patterns.

We compared the spectral participation ratio to the spreading behavior across coin parameters and noise levels, and found consistent correlation in both regimes. 
In the walk-on-the-line regime, low participation ratios predict sub-diffusive spread, while high ratios align with ballistic or super-diffusive evolution. 
In the walk-on-the-cycle regime, low participation ratios predict low saturation level of the mean squared displacement, reflecting localization, while high ratios align with high saturation, reflecting delocalization.

Our results show that coin structure, disorder strength, and cyclic topology jointly determine the dynamical evolution of quantum walks. 
The participation ratio offers a fast static diagnostic that complements dynamical simulations and may be useful in analyzing noisy quantum devices.

Future work may address the role of dynamic disorder, exploring time-dependent perturbations and their consequences for localization and quantum transport \cite{Rebentrost2009p033003, Hideaki2011p195139, Bardarson2012p017202, Asch2012p194205, Agarwal2015p160401, Lin2022p3229, Wang2018p11361140}.


\section*{Funding}
Not Applicable

\section*{Acknowledgment}
B.~M.~R.~L. acknowledge fruitful discussion with Emiliano Etienne Iracheta.

\section*{Disclosures} 
The authors declare no conflicts of interest.

\section*{Data availability} 
Data underlying the results presented in this paper may be obtained from the authors upon reasonable request.


%

\end{document}